\RequirePackage{rotating}
\documentclass[10pt, conference]{IEEEtran}

\usepackage{amsmath}
\usepackage[ruled,vlined,linesnumbered]{algorithm2e}
\usepackage{amsmath}
\usepackage[T1]{fontenc}
\usepackage{graphicx}
\usepackage{textcomp}
\usepackage{tikz}
\usepackage{hyperref}
\usepackage{textcomp}
\usepackage{color}
\usepackage[utf8]{inputenc}
\usepackage{fancyvrb}
\usepackage{xcolor}
\usepackage{colortbl}
\usepackage{hhline}
\usepackage{lscape}
\usepackage{upquote}
\usepackage{array}
\usepackage{adjustbox}
\usepackage{rotating}
\usepackage{stmaryrd}
\usepackage{pifont}
\usepackage{multirow}
\usepackage{wrapfig}
\usepackage{verbatim}
\usepackage{colortbl}
\usepackage{rotating}

\newcolumntype{L}[1]{>{\raggedright\arraybackslash}m{#1}}
\newcolumntype{C}[1]{>{\centering\arraybackslash}m{#1}}
\newcolumntype{R}[1]{>{\raggedleft\arraybackslash}m{#1}}

\fvset{commandchars=\\\{\}}

\definecolor{Green}{rgb}{0.0, 0.56, 0.0}
\definecolor{Gray}{gray}{0.85}
\definecolor{LightBlue}{cmyk}{0.51, 0.1, 0, 0}
\definecolor{Maroon}{cmyk}{0.32, 1.0, 1.0, .46}

\newcommand{\LightBlue}[1]{\textcolor{LightBlue}{\textbf{#1}}}
\newcommand{\Blue}[1]{\textcolor{blue}{\textbf{#1}}}
\newcommand{\Green}[1]{\textcolor{Green}{\textbf{#1}}}
\newcommand{\Red}[1]{\textcolor{red}{\textbf{#1}}}

\newcommand{\CodeIn}[1]{\begin{small}\texttt{#1}\end{small}}
\newcommand{\Comment}[1]{}
\newcommand{\NoType}[1]{} 
\newcommand{\Space}[1]{}

\newcommand{\DefMacro}[2]{\expandafter\newcommand\csname rmk-#1\endcsname{#2}}
\newcommand{\UseMacro}[1]{\csname rmk-#1\endcsname}

\DefMacro{check}{\ding{51}}
\DefMacro{corr}{\ding{72}}
\DefMacro{plausible}{\ding{51}}

\newcommand{\tool}{\emph{Crucible}}

\newcommand{\ACaret}{\raisebox{-0.6ex}{\textasciicircum}}
\newcommand{\AStar}{*}

\usepackage{graphicx}
\begin{document}

\title{Crucible: Graphical Test Cases for Alloy Models}


\author{
\IEEEauthorblockN{Adam G. Emerson}
\IEEEauthorblockA{
\textit{University of Texas at Arlington} \\
Arlington, TX USA \\
adam.emerson@mavs.uta.edu}
\and
\IEEEauthorblockN{Allison Sullivan}
\IEEEauthorblockA{
\textit{University of Texas at Arlington} \\
Arlington, TX USA \\
allison.sullivan@uta.edu}
}

%
%


%
%

%
%
\maketitle   

\begin{abstract}
Alloy is a declarative modeling language that is well suited for verifying system designs. Alloy models are automatically analyzed using the Analyzer, a toolset that helps the user understand their system by displaying the consequences of their properties, helping identify any missing or incorrect properties, and exploring the impact of modifications to those properties. To achieve this, the Analyzer invokes off-the-shelf SAT solvers to search for scenarios, which are assignments to the sets and relations of the model such that all executed formulas hold. To help write more accurate software models, Alloy has a unit testing framework, AUnit, which allows users to outline specific scenarios and check if those scenarios are correctly generated or prevented by their model. Unfortunately, AUnit currently only supports textual specifications of scenarios. This paper introduces Crucible, which allows users to graphically create AUnit test cases. In addition, Crucible provides automated guidance to users to ensure they are creating well structured, valuable test cases. As a result, Crucible eases the burden of adopting AUnit and brings AUnit test case creation more in line with how Alloy scenarios are commonly interacted with, which is graphically.
\end{abstract}

\section{Introduction}
In today's society, we are becoming increasingly dependent on software systems. However, we also constantly witness the negative impacts of buggy software. One way to help develop better software systems is to leverage software models. When forming requirements, software models can be used to clearly communicate to all stakeholders both the desired system as well as the environment it will be deployed in. When creating designs and implementations, software models can help reason over how well the design and implementation choices satisfy the requirements. As such, software models can help detect flaws earlier in development and thus aid in the delivery of more reliable systems.

Alloy~\cite{JacksonAlloyBook2006} is a relational modeling language. A key strength of Alloy is the ability to develop models in the Analyzer, an automatic analysis engine based on off-the-shelf SAT solvers, which the Analyzer uses to generate scenarios that highlight how the modeled properties either hold or are refuted, as desired. The user is able to iterate over these scenarios one by one, inspecting them for correctness. 
Alloy has been used to verify software system designs~\cite{Zave15,BagheriETAL2018,WickersonETAL2017,ChongETAL2018},
and to perform various forms of analyses over the corresponding implementation, including deep static checking~\cite{JacksonVaziri00Bugs,TACOGaleottiETALTSE2013}, systematic testing~\cite{MarinovKhurshid01TestEra}, data structure repair~\cite{ZaeemKhurshidECOOP2010}, automated debugging~\cite{GopinathETALTACAS2011} and to synthesize security attacks~\cite{websecurity,Margrave,CheckMateMicro2019}. 

However, to gain the many benefits that come from utilizing software models, the model itself needs to be correct. Unfortunately, while Alloy offers succinct formulation of complex properties, Alloy's support for expressive operators, such as transitive closure and quantified formulas, can make writing non-trivial properties challenging, especially for beginner users. In Alloy, there are two types of faults that can appear in a model: (1)~\emph{under-constrained} faults in which the model allows scenarios it should prevent, and (2)~\emph{over-constrained} faults in which the model prevents scenarios it should allow.
To help detect these types of faults in an Alloy model, a unit testing framework, AUnit, was created~\cite{AUnitConcept,AUnit}. AUnit enables users to outline a specific scenario they expect their model to allow or prevent and then check that this behavior actually occurs. This improves upon the previous ad-hoc practices that require users to either (1)  enumerate scenarios until finding one that is malformed or (2) enumerate all scenarios and realize one was missing, in order to determine if their model is faulty. 

\usetikzlibrary{shapes.geometric, arrows}
\tikzstyle{arrow} = [line width=1.5pt,->,>=stealth]
\tikzstyle{darrow} = [line width=1.5pt,<->,>=stealth]
\tikzstyle{NodeAtom} = [rectangle, minimum width=.75cm, minimum height=.75cm, text centered, draw=black, fill=yellow!75!orange]
\tikzstyle{ListAtom} = [rectangle, minimum width=.75cm, minimum height=.75cm, text centered, draw=black, fill=orange!75!yellow]

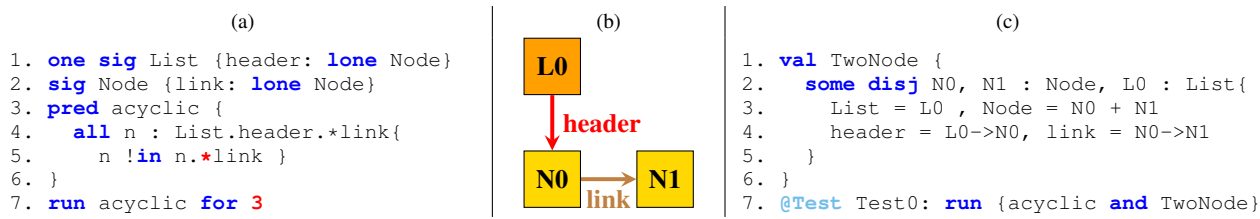
\begin{figure*}
    \centering

\begin{tabular}[t]{c|c|c}
\footnotesize (a) & \footnotesize (b) & \footnotesize (c)  \\
\begin{minipage}[t]{.7\columnwidth}
\footnotesize
\begin{Verbatim}[]
1. \Blue{one sig} List \{header: \Blue{lone} Node\}
2. \Blue{sig} Node \{link: \Blue{lone} Node\} 
3. \Blue{pred} acyclic \{
4.   \Blue{all} n : List.header.*link\{
5.     n !\Blue{in} n.\Red{*}link \}
6. \}
7. \Blue{run} acyclic \Blue{for} \Red{3}
\end{Verbatim}

\end{minipage}

&

\begin{minipage}[t]{.3\columnwidth}
\begin{center}
\begin{tikzpicture}[baseline,node distance=1.5cm]
\node (L0) [ListAtom] {\textbf{L0}};
\node (N0) [NodeAtom, below of=L0] {\textbf{N0}};
\node (N1) [NodeAtom, right of=N0] {\textbf{N1}};
\begin{scope}[red]
\draw [arrow] (L0) -- node[anchor=west] {\textbf{header}} (N0);
\end{scope}[red]
\begin{scope}[brown]
\draw [arrow] (N0) -- node[anchor=north] {\textbf{link}} (N1);
\end{scope}[brown]
\end{tikzpicture}
\end{center}
\end{minipage}

&

\begin{minipage}[t]{.8\columnwidth}
\footnotesize
\begin{Verbatim}[]
1. \Blue{val} TwoNode \{
2.   \Blue{some disj} N0, N1 : Node, L0 : List\{
3.     List = L0 , Node = N0 + N1
4.     header = L0->N0, link = N0->N1
5.   \}
6. \} 
7. \LightBlue{@Test} Test0: \Blue{run} \{acyclic \Blue{and} TwoNode\}
\end{Verbatim}
\end{minipage}  \\

\end{tabular}
    \caption{Faulty Alloy Model and Fault-Revealing AUnit Test Case}
    \label{fig:testcase}
\end{figure*}

AUnit laid the foundation to bring a number of proven imperative testing practices to Alloy, including mutation testing~\cite{MuAlloyTool}, automated test generation~\cite{AUnit}, fault localization~\cite{AlloyFL}, automated repair~\cite{ARepair} and partial synthesis of models~\cite{ASketchABZ}. These extensions help establish a comprehensive testing environment for Alloy  that is similar to the robust testing support that imperative languages like Java have. 
Unfortunately, when it comes to the actual creation of an AUnit test case, the user is required to outline the valuation portion of a test case textually as a series of set equality statements that are wrapped around an existentially quantified formula.

However, valuations, which outline scenarios, are commonly interacted with graphically not textually. This results in a gap between the user's mental model of an Alloy valuation and the way the user currently has to create the valuation in order for the Analyzer to successfully reproduce it and perform unit testing. This paper addresses this issue by introducing \tool{}, which establishes support for users to graphically create test cases. In addition, \tool{} leverages the underlying model to help guide the user to create well-formed test cases by warning users when they attempt to create test cases that violate the model's structural constraints, which helps the user create stronger, more effective test suites.

In this paper, we make the following contributions:

\newenvironment{Contributions}{}{} 
\newcommand{\Contribution}[1]{\noindent\textbf{#1:}} 

\begin{Contributions}
\Contribution{Graphical Specification of Test Cases} We introduce \tool{}, which enables users to create AUnit test cases graphically through a drag and drop interface. 

\Contribution{Automated Guidance} Based on the underlying model, \tool{} will prevent users from creating malformed test cases and inform the user of which part of the model's signatures prevents the test case from being feasible.

\Contribution{Evaluation} We evaluate the overhead of \tool's translation between the graphical and textual representation of a test case. We also use key examples to highlight the need to support the graphical creation of test cases to ease the burden of creating test cases for models with complicated structures.

\Contribution{Open Source} We release \tool{} as an open-source toolset at: \url{https://github.com/Crucible-Alloy/Crucible}.

\end{Contributions}
\section{Background}\label{sec:example}

\subsection{Alloy}
To highlight how modeling in Alloy works, Figure~\ref{fig:testcase}~(a) depicts a faulty model of a singly-linked list with an acyclic constraint. Signature paragraphs and the relations declared within introduce atoms and their relationships (lines 1 - 2). Line 1 introduces a named set \CodeIn{List} and uses the relation \CodeIn{header} to express that each \CodeIn{List} atom has zero or one header nodes (\CodeIn{lone}). Similarly, line 2 introduces the named set \CodeIn{Node} and uses the \CodeIn{link} relation to express that each \CodeIn{Node} atom points to zero or one other nodes.  Predicate paragraphs introduce named formulas that can be invoked elsewhere (lines 3 - 6). The predicate \CodeIn{Acyclic} uses universal quantification (\CodeIn{all}), set exclusion (\CodeIn{!in}), relation join (\CodeIn{.}),  and reflexive transitive closure (\CodeIn{\AStar}) to try to express the idea that ``for all nodes in the list, no node is reachable from themselves following one or more traversals down the link relation.'' The fault, in red, can be corrected by replacing reflexive transitive closure~(`\CodeIn{\AStar}') with transitive closure~(`\CodeIn{\ACaret}'), which will produce a set that will not include the node itself. 

Commands indicate which formulas to invoke and what scope to explore.  The command on line 7 asks the Analyzer to search for satisfying assignments to all the sets of the model (\CodeIn{List}, \CodeIn{header}, \CodeIn{Node}, and \CodeIn{link}) such that \CodeIn{Acyclic} is true using up to 3 \CodeIn{List} atoms and 3 \CodeIn{Node} atoms. A user can iterate over all the scenarios found by the SAT solver one by one. At a conceptual level, each scenario depicts behavior currently allowed by the modeled system. Figure~\ref{fig:testcase} (b) graphically displays a scenario that user would expect to be found by the Analyzer when the command at line 7 is executed: a list with two nodes and no cycles. Therefore, the user would expect to encounter this scenario at some point. However, due to the fault, this will never happen.

\subsection{AUnit}
AUnit addresses the need to have a systematic method to check the correctness of Alloy models~\cite{AUnitConcept}. Before AUnit, there was no formal notion of ``testing'' in the Analyzer. As a result, experienced users would employ a range of ad-hoc techniques, such as enumerating all scenarios -- which can number in the thousands -- and visually inspecting them for issues, a process which is both time consuming and error prone. 
Moreover, the number of scenarios can be in the hundreds and the order scenarios are presented in based on the order the backend SAT solver finds them, which means the scenarios are effectively unordered. Altogether, this makes the enumeration-inspection process not practical. For instance, since scenarios are unordered, the user cannot count on relying them in order of increasing size, which would make it more feasible to catch a missing scenario.   

The key insight behind AUnit is that unit testing, the most effective way to validate \emph{code}, provides a blueprint on how to validate \emph{models}. 
Specifically, an AUnit test case consists of two components: a \emph{valuation}, which is an assignment to the sets and relations of the model, and a \emph{command}, which specifies the Alloy formulas under test. A test case passes if the valuation is a valid scenario of the associated command; otherwise, the test fails. AUnit enables a user to directly ensure a specific scenario they have in mind is correctly generated -- which checks for over-constrained faults -- or prevented -- which checks for under-constrained faults --  without having to rely on encountering the scenario, or not, in the Analyzer. 

To demonstrate,  Figure~\ref{fig:testcase}~(b) and (c) depicts an AUnit test case graphically and textually which reveals the faulty behavior. This test case is based on a scenario which is valid for the correct model, but is incorrectly invalid for the faulty model. The valuation, outlined in lines 1-6 in Figure~\ref{fig:testcase}~(c), assigns all the sets  (\CodeIn{List} and \CodeIn{Node}) and relations (\CodeIn{header} and \CodeIn{link}) of the singly-linked list to concrete values, creating a single scenario to reason over, by using existential quantification (`\CodeIn{some}') and the disjoint operator (`\CodeIn{disj}') to declare local variables and set equality (`\CodeIn{=}') to assign these local variables to constrain the sets of the model. The command given in line 7 in Figure~\ref{fig:testcase}~(c) outlines the \CodeIn{Acyclic} predicate as the formula under test. When the Analyzer executes the test case, the command is unexpectedly unsatisfiable, revealing the fault. Prior work has extended the Analyzer with the ability to natively declare AUnit test cases by extending the grammar with new keywords: \CodeIn{val} to outlined valuations and \CodeIn{@Test} to flag which Alloy commands refer to test executions~\cite{AUnitTool}.

\subsection{Challenge: Specifying Test Cases Textually}
The textual format, as seen in Figure~\ref{fig:testcase} (c), can be tedious to provide, especially if the test case reasons over multiple states, contains numerous atoms or has higher arity relations. These features in a model can quickly bloat the length of the textual representation, impacting the readability, and thus usability, of an AUnit test. For instance, if a model has 10 signatures, then the atoms for all 10 signatures need to be accounted for as local variables declared by the existentially quantified formula. Then, the user needs to make a set equality formula for each signature. For a scope of 3, this would be declaring up to 30 variables and creating 10 separate set equality formulas for just the signatures. The test case would still need to create set equality formulas to account for any relations, which could utilize all 30 variables multiple times.

The textual representation is also not in line with the default way users inspect scenarios, which is graphically. As a result, requiring users to supply the textual representation increases the burden on the user to accurately translate their mental image of a scenario into a valid test case. In fact, recent work has demonstrated the importance of spatial cognition ability in solving Alloy tasks for both novice and expert users~\cite{alloyformalise23study}. This is especially true for writing AUnit test cases, which requires the user to mentally picture a scenario of interest and then accurately write constraints to, in turn, generate that exact scenario. \tool{} addresses these pain points by allowing the user to directly supply the graphical representation of a scenario, which \tool{} will automatically translate to the textual representation.

\section{\tool{}}\label{sec:open}

This section outlines important implementation details of \tool, which is a standalone desktop application. We first present an overview of \tool's system architecture. Then we step over the process of creating test cases, including how \tool{} ensures a user cannot create a malformed test case and how \tool{} automatically translates graphical renderings to textual test cases.

\subsection{Framework Overview}
\begin{figure}
    \centering
    \includegraphics[width=0.9\columnwidth]{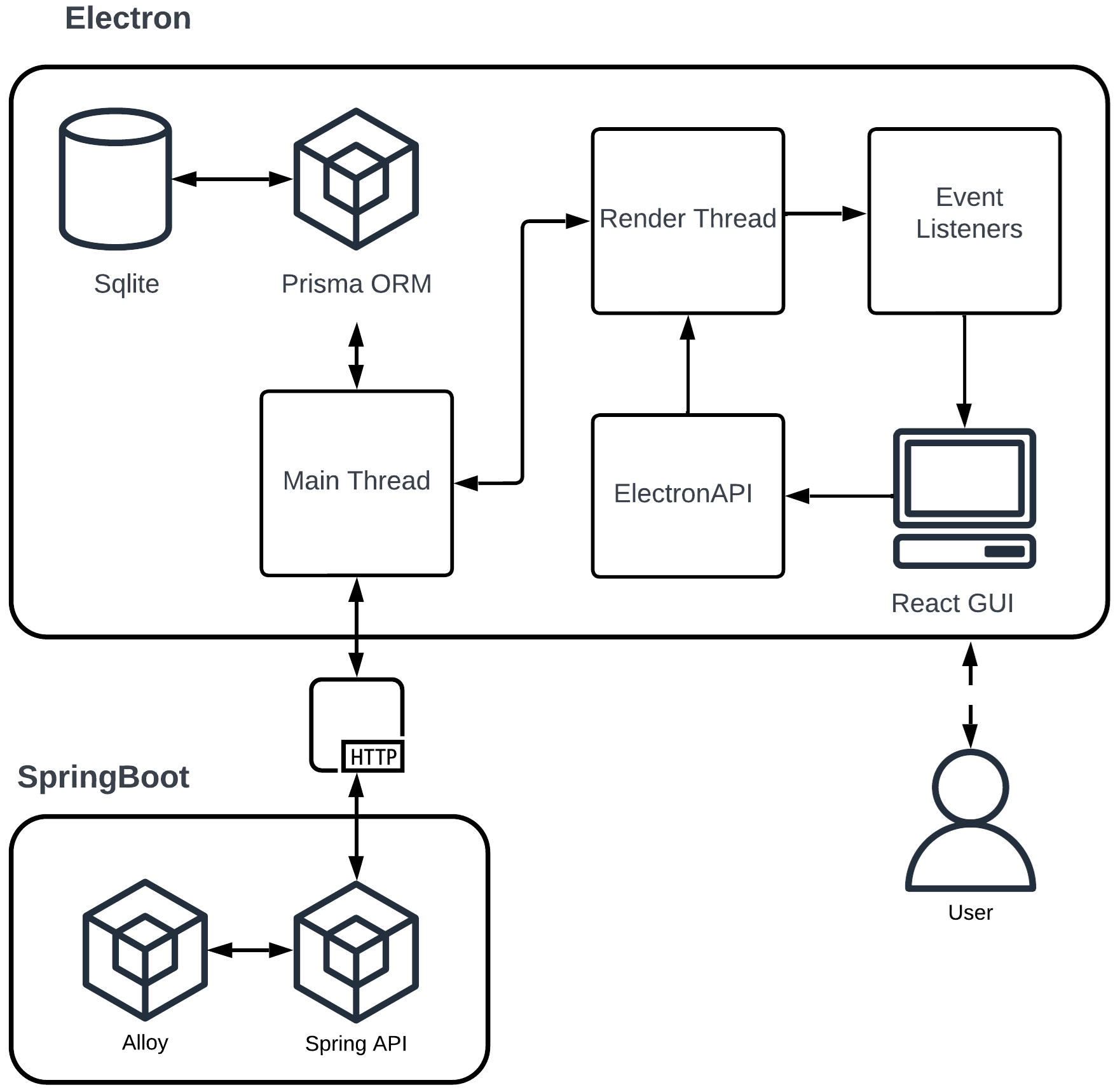}
    \caption{\tool{} Software Architecture Diagram}
    \label{fig:sa}
\end{figure}

Figure~\ref{fig:sa} displays the high level software architecture for \tool. \tool{} looks to connect two main processes together: (1) the React graphical user interface (GUI), which the user will use to create AUnit test cases and (2) the Alloy Analyzer, which will execute the AUnit test cases. While the Analyzer is written in Java, our GUI is built using popular web technologies; Typescript, React, and Electron. To ensure the two processes are able to communicate, the Analyzer is wrapped in a SpringBoot REST API which is launched in conjunction with the GUI. This API handles the communication between processes when Alloy is needed, which occurs (1) when the model is initially parsed and its signature and predicate information is stored for use by \tool{} and (2) when a test case is executed. Although it may have been a more obvious choice to write \tool{} using a Java framework like JavaFX \cite{javafx}, which would do away with the additional overhead of an API, we opted for a web-based desktop platform for several reasons. 

Firstly, the core purpose of \tool{} is to improve both user experience and efficiency while writing AUnit test cases. With this in mind, it is critical that the application's interface is familiar, intuitive, and performant. Front-end frameworks like React, Angular, and Vue, have all been driving forces in UI development over the last decade~\cite{saks2019javascript}. These frameworks, particularly React, have grown to support a massive ecosystem of well-documented off-the-shelf primitive components that serve to expedite development of web and desktop applications alike. Thanks to widespread adoption of React and the GUI oriented nature of the web, the available open-source components are of a higher-quality than those which may be available within the Java ecosystem.

\begin{figure*}
    \centering
    \begin{tabular}{c|c}
       \includegraphics[width=0.9\columnwidth]{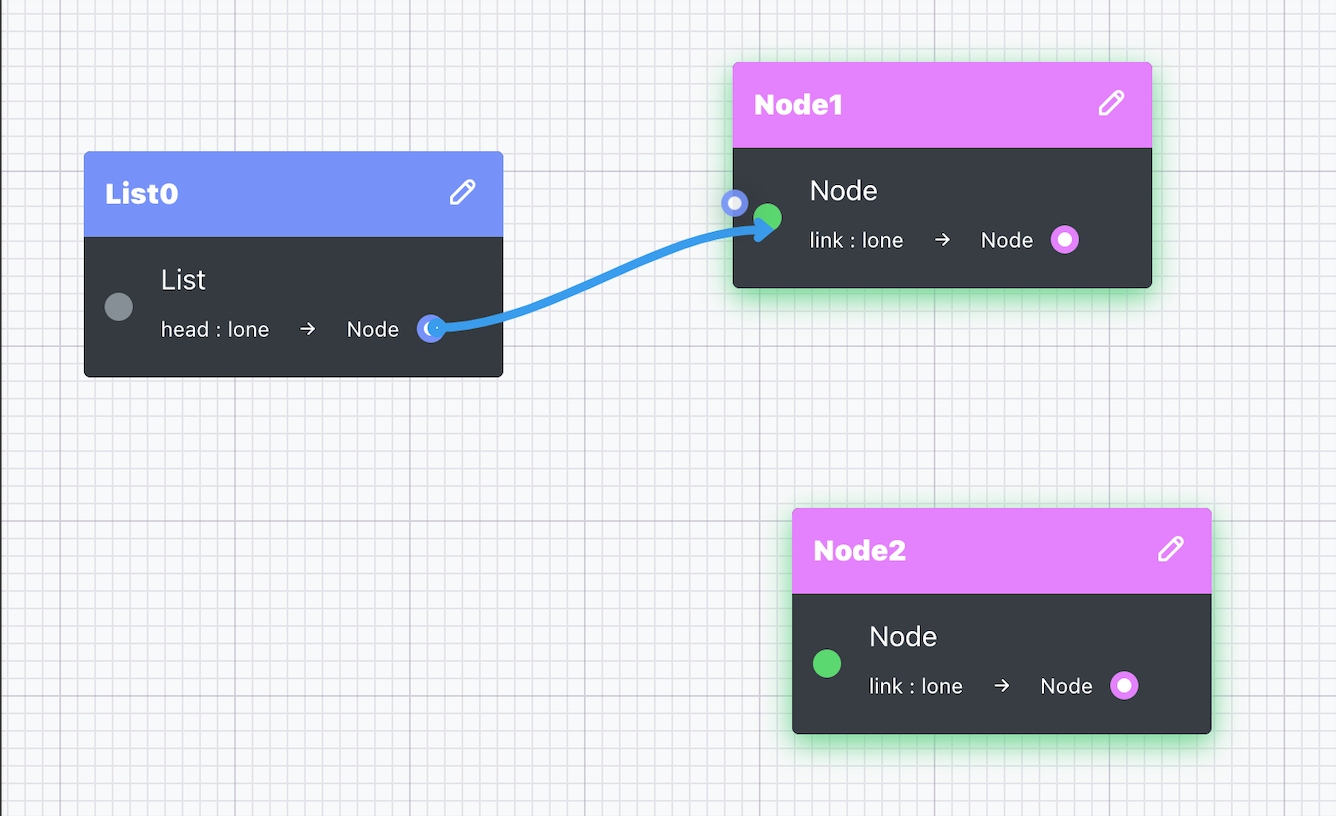}  & \includegraphics[width=0.9\columnwidth]{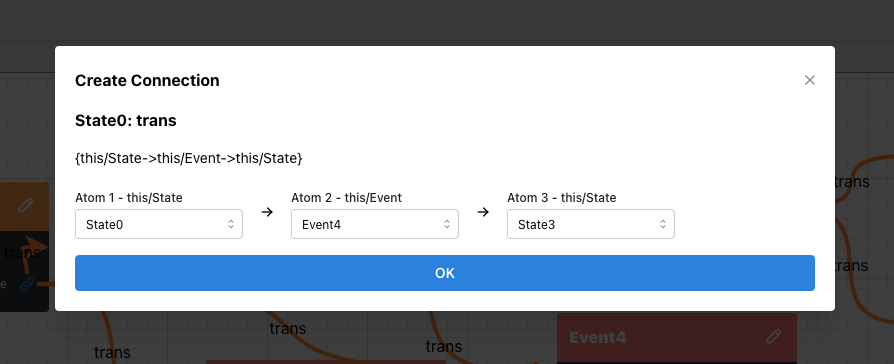} \\
      \footnotesize (a)  & \footnotesize (b)
    \end{tabular}
    
    \caption{\tool{} canvas in action: (a) highlighting allowed connections and (b) declaring higher arity connections.}
    \label{fig:screen-connections}
\end{figure*}

Secondly, of equal importance is the portability and accessibility of \tool. By writing \tool{} with web technologies, specifically React and Typescript, we ensure that deployment of \tool{} is even more flexible than that afforded by the JVM. If in the future a third-party wished to host \tool{} on a server for remote learning or some other application, it would be relatively trivial to port the codebase into a fully fledged web application. Notably, one of the main educational environments for Alloy is the Alloy4Fun website, which provides an online hosted platform for editing, sharing and interpreting Alloy models~\cite{alloy4fun}. 
\tool's current form factor can more easily be integrated with Alloy4Fun than a JavaFX variant of \tool. As a result, Alloy4Fun can feasibly be updated to give new users who are exploring Alloy for the first time a native environment to easily address ``how do I test my Alloy program?'' 

\subsection{Creating a Project}
To get started with \tool, the user first needs to input their Alloy model into a new "Project".  Each project consists of a single Alloy file, and acts as an organizational object for all of the tests written for the model. Upon selecting an Alloy file, the file path is sent via the API to the Java process, where it is passed into the Alloy Analyzer and converted into a JSON object of the model's signature and predicate definitions. The JSON object is then returned to the GUI process where it is cached in a SQLite database for subsequent usage. The SQLite database is used by \tool{} to help provide automated guidance, as it contains all the information needed to know any structural constraints attached to a model's signatures, such as multiplicity constraints, and any defined relations. The SQLite object also stores all the information needed such that \tool{} can easily allow users to test any predicate currently defined in the model, including ensuring that the user provides the required parameters for a predicate, if needed.

\subsection{Creating a Test Case}
Once the model has been imported and a project has been initialized, the user can then create any number of test cases. Each test is uniquely named, and consists of (1) a \emph{canvas} onto which the user can spawn any number of atoms and connections, as allowed by the model, and (2) a set of all predicates and assertions declared in the model, by which the user can select which predicate(s) to consider under evaluation. Atoms are elements of signature sets and connections populate relations onto the canvas, e.g. to replicate the valuation in Figure~\ref{fig:testcase} (b), \CodeIn{N0} would be a placeable \CodeIn{Node} atom and the red \CodeIn{header} directed line would be a placeable connection. 

In \tool's project view, the model's signatures are presented as small tokens in a drawer menu on the left hand side. Each token displays the signature's name, relations, and multiplicity. Every token is assigned a color upon project initialization which can be edited to the user's liking. To build a test, atoms are dragged from the appropriate signature token and onto the canvas. Upon being added to the canvas, each atom is automatically given a unique nickname for use in command string generation and to help identify the atom as a predicate parameter. Once a sufficient number of atoms have been added to the canvas, connections can be made between them through a similar drag and drop interaction, or in the case of connections with a higher arity, a modal pop-up. Changes to a test case are saved automatically as they are made, further streamlining the process and allowing a user to focus on the task at hand.

\subsection{Automated Guidance}
As atoms are dropped onto the canvas, \tool{} checks the current canvas state for multiplicity violations and alerts the user if they are attempting an addition that violates the model. Consider the \CodeIn{List} signature from our singly-linked list model:

\begin{footnotesize}
\begin{Verbatim}[frame=lines,rulecolor=\color{lightgray}]
1. \Red{one} \Blue{sig} List \{header: \Blue{lone} Node\}
\end{Verbatim}
\end{footnotesize}

\noindent This signature uses the singleton multiplicity constraint, meaning that there can only be one \CodeIn{List} object for any valid scenario of this model. Therefore, if a user tries to form a test case with multiple \CodeIn{List} objects, \tool{} will alert the user that she is attempting to violate the structural constraints of the \CodeIn{List} signature. This is an important detail for the user to be aware of, as any test with more than one \CodeIn{List} object will \textit{always} be prevented by the model. Therefore, \tool's proactive guidance ensures the user will not incorrectly draw conclusions about the correctness of any predicates under test. 
To illustrate, if the user built the following valuation:

\begin{center}
\begin{tikzpicture}[baseline,node distance=1.6cm]
\node (L0) [ListAtom] {\textbf{L1}};
\node (L1) [ListAtom, left of=L0] {\textbf{L0}};
\node (N0) [NodeAtom, below of=L0] {\textbf{N0}};
\node (N1) [NodeAtom, right of=N0] {\textbf{N1}};
\begin{scope}[red]
\draw [arrow] (L0) -- node[anchor=west] {\textbf{header}} (N0);
\end{scope}[red]
\begin{scope}[brown]
\draw [darrow] (N0) -- node[anchor=north] {\textbf{link}} (N1);
\end{scope}[brown]
\end{tikzpicture}
\end{center}

\noindent and checked that it is successfully prevented by the \CodeIn{acyclic} predicate, the user could build a false sense of security in the accuracy of their \CodeIn{acyclic} predicate, as the valuation will always be prevented due to the presence of two \CodeIn{List} atoms, regardless of any formulas in \CodeIn{acyclic}.

As a user initiates a connection interaction to add relations to the canvas, valid connections targets will be highlighted based on the defined relations in the model, as seen in Figure~\ref{fig:screen-connections}. If a user attempts to make a connection to a non-valid target, they will be notified of the issue. In addition, even if the user has the right target, if the user attempts to make a connection that violates a relational multiplicity constraint, they will again be notified and the action will be prevented. 


 To illustrate, consider the \CodeIn{header} relation:

\begin{footnotesize}
\begin{Verbatim}[frame=lines,rulecolor=\color{lightgray}]
1. \Blue{one} \Blue{sig} List \{header: \Red{lone} Node\}
\end{Verbatim}
\end{footnotesize}

\noindent The \CodeIn{header} relation conveys two important pieces of information. First, the \CodeIn{header} relation is meant to connect a \CodeIn{List} atom to a \CodeIn{Node} atom. Second, the multiplicity constraint \CodeIn{lone} further restricts this by asserting that for each \CodeIn{List} atom, the \CodeIn{header} relation can only connect that \CodeIn{List} atom to either no \CodeIn{Node} atom or exactly one \CodeIn{Node} atom. As seen in Figure~\ref{fig:screen-connections} (a), when a user wants to add a \CodeIn{header} relation to their test case, the user will see that the relation must start on a \CodeIn{List} object, and only end connections on \CodeIn{Node} atoms will be highlighted.

For higher arity (3+) relations, we do not currently enforce multiplicity constraints. However, if a higher arity connection is specified of the form ``\CodeIn{a->b->c}'' and the user deletes the connection ``\CodeIn{a->b},'' we automatically remove ``\CodeIn{b->c}'' from the canvas. In addition, to add a higher arity connection, we created a tailored modal that helps guide the user to specify each segment of the connection with drop-down menus that populate with only the valid atom options. Figure~\ref{fig:screen-connections} (b) shows the higher arity relation modal. Since our singly linked list model only has binary relations, we use the LTS from our evaluation in Section~\ref{sec:casestudy} to highlight this interface and its corresponding guidance.

\tool's proactive nature of preventing users from creating test cases that violate the constraints outlined in signature paragraphs, and alerting users as to why what they are attempting to create is malformed, ensures that user is both aware of how the structural constraints of their model restrict the shape of valid valuations and ensures the user knows that the valuation is prevented because of these structural constraints, regardless of any command portion the user may have placed on the test case. This directly prevents the false sense of security that can be formed about a predicate mentioned earlier, where creating a test case with more than one list does not help us evaluate the \CodeIn{acyclic} predicate or \emph{any} other system property the user writes. 

As a tradeoff for the guidance we provide, users cannot directly form test cases for constraints enforced by the signature paragraphs of their model. 
We do believe that users should ensure their signature paragraphs are correct and modify them if they are not. 
Since \tool{} proactively gives the user detailed error message pop-ups when the user tries to violate multiplicity constraints and grays out improper relation connections, the user still interactively explores these constraints within \tool, enabling the user to still check the accuracy of their signature paragraphs, albeit indirectly. However, we feel the tradeoff is worthwhile to ensure users are actually testing the predicates they intend to. 
 

\subsection{Automated Translation}\label{sec:encoding}
Running a test in \tool{} is as simple as pressing a button. At runtime, the test's canvas is converted into an AUnit command string that the Alloy API can process and execute using the Analyzer. The command string is a series of valid Alloy formulas that, when executed, will produce just the scenario outlined on the canvas. To create the command string, \tool{} processes each atom captured on the canvas, which includes tying the atom to its unique nickname 
and capturing all of the declared connections attached to this atom. 
Then, \tool{} builds a mapping from each atom to the atom's associated signature. Once this mapping is formed, for each signature, \tool{} generates an existentially quantified formula of the form:

\begin{footnotesize}
\begin{Verbatim}[frame=lines,rulecolor=\color{lightgray}]
some disj [nickname]* : [signature name] \{ 
\end{Verbatim}
\end{footnotesize}

The \CodeIn{disj} keywords ensures that each variable name listed will produce a distinct atom for any satisfying instance. For example, in Figure~\ref{fig:screen-connections} the following will get generated based on the state of the canvas:

\begin{footnotesize}
\begin{Verbatim}[frame=lines,rulecolor=\color{lightgray}]
\Red{some disj} L0 : List \{
\Red{some disj} N0, N1 : Node \{ 
\end{Verbatim}
\end{footnotesize}

\noindent where \CodeIn{N0} and \CodeIn{N1} cannot be represented by the same atom for any scenario produced by the Analyzer. \tool{} processes each signature in the order they are declared in the model. Once all nicknames have been declared as local variables, \tool{} generates a set equality formula of the form: 

\begin{footnotesize}
\begin{Verbatim}[frame=lines,rulecolor=\color{lightgray}]
[signature name] = [nickname] (+ [nickname])* |
                    no [signaure name]
 [relation name] = [connection] (+ [connection])* |
                    no [relation name]
\end{Verbatim}
\end{footnotesize}

\noindent  where (\CodeIn{+}) is set union. As a result, the value each signature set can take for any generated scenario is restricted to just the declared local variables of that type and nothing else. In addition, relations are restricted to the connections specified by the atoms. Likewise, the set equality formula must be declared within the scope of the local variables. If there are no atoms in the canvas for a signature or no connections for a relation, then the empty set operator (\CodeIn{no}) is used instead to ensure that this signature or relation does not appear in the corresponding scenario the Analyzer generates to satisfying the outlined valuation. For our example this will result in the following:

\begin{footnotesize}
\begin{Verbatim}[frame=lines,rulecolor=\color{lightgray}]
\Blue{some disj} L0 : List \{  \Green{//Start of L0 scope}
\Blue{some disj} N0, N1 : Node \{ \Green{//Start of N0, N1 scope}
  List \Red{=} L0,        Node \Red{=} N0 + N1
  header \Red{=} L0->N0, linke \Red{=} N0-N1
\end{Verbatim}
\end{footnotesize}

\begin{figure}
    \centering
    \includegraphics[width=0.9\columnwidth]{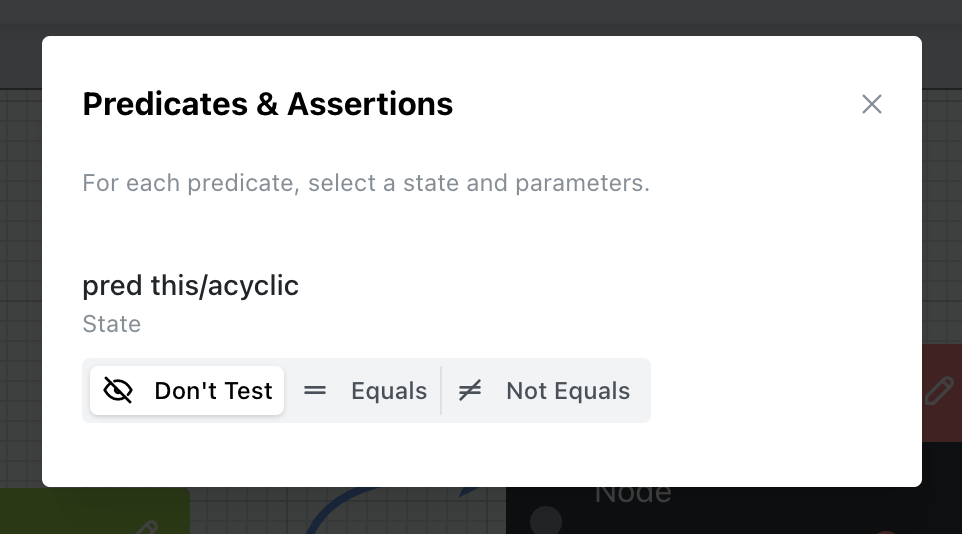}
    \caption{\tool{} Predicates Modal}
    \label{fig:pred-gui}
\end{figure}

Before running a test the user has the option of modifying the predicate(s) they wish to test.  Users can adjust predicates by opening the predicate modal, where they will be able to assign atoms by nickname as parameters and chose one of the states for the predicate, as seen in Figure~\ref{fig:pred-gui}. The states are "Don't Test" (null), where the predicate is not tested, "Valid", where the valuation is expected to be generated by the predicate, and "Invalid" where the valuation is expected to be prevented by the predicate. Based on the user's selection, \tool{} will append the following information to the command string:

\begin{footnotesize}
\begin{Verbatim}[frame=lines,rulecolor=\color{lightgray}]
null: ""
true: [predicate name][(param)*]
false: ![predicate name][(param)*]
\end{Verbatim}
\end{footnotesize}

For our example, this will result in the following:
\begin{footnotesize}
\begin{Verbatim}[frame=lines,rulecolor=\color{lightgray}]
\Blue{some disj} L0 : List \{  \Green{//Start of L0 scope}
\Blue{some disj} N0, N1 : Node \{ \Green{//Start of N0, N1 scope}
  List = L0,       Node = N0 + N1
  header = L0->N0, linke = N0-N1
  \Red{acyclic[]}
\}\}
\end{Verbatim}
\end{footnotesize}

\noindent Outlining the predicate under test within the scope of the local variables is important, as predicates can have parameters. For instance, if acyclic was defined as ``\CodeIn{acyclic[l : List]}'' then the predicate under test would become ``\CodeIn{acyclic[L0]},'' which the Analyzer will fail to compile if the predicate call is located outside the scope of the \CodeIn{L0} variable. After all predicates under test are specified, the local variable scopes can be closed.

Once the command string is generated, the string is sent to the Java Process to be run in the Analyzer. If the command string is satisfiable, the API returns a success code to the GUI Process, and the user is notified that the test has passed. Alternatively, if the test is not satisfiable, the API returns the failure.  It is worth noting that due to the multiplicity constraints being strictly enforced on the canvas, it is not possible to generate a failing test case without enabling one or more predicates, with the exception of higher arity relations which require further exploration.

\section{Evaluation}\label{sec:eval}
We evaluate \tool{} in two ways. First, we evaluate the overhead of translating graphical renderings into executable test cases. Second, we conduct an illustrative case study over a select set of models to highlight how \tool{} can ease the burden of creating test cases for models with tedious features for textual test case creation.

\subsection{Overhead} 
Table~\ref{tab:overhead} shows the runtime to translate the canvas state of \tool{} into a command string for increasingly larger and larger test cases. Column \textbf{Model} conveys the model under evaluation. The next two columns outline the size of the model: column \textbf{\#Sig} is the number of signatures and column \textbf{\#Rel} is the number of relations in the model. To convey the size of the test case, column \textbf{\#Atoms} displays the number of atoms on the canvas and column \textbf{\#Con} shows the number of connections on the canvas. Column \textbf{Time[ms]} conveys the average time it takes (rounded to the nearest millisecond) across ten executions for \tool{} to generate the command string once the execute test button in pressed. To perform the calculations, we create incrementally larger graphical test cases in \tool{} for the two models we explore in our case study: the LTS model, which contains a higher arity relation, and the CV model, which contains a large number of signature and relations.

\begin{table}[]
    \centering
    \caption{Overhead of Translation}
    \label{tab:overhead}
    \begin{tabular}{rrr|rrr}
\hline
\textbf{Model} & \textbf{\#Sig} & \textbf{\#Rel} & \textbf{\#Atoms} & \textbf{\#Con} & \textbf{Time{[}ms{]}} \\ \hline
\multicolumn{1}{l}{\multirow{5}{*}{{LTS}}} & \multicolumn{1}{l}{\multirow{5}{*}{3}} & \multicolumn{1}{l|}{\multirow{5}{*}{1}} & \cellcolor[HTML]{EFEFEF} 3 & \cellcolor[HTML]{EFEFEF} 3 & 6\cellcolor[HTML]{EFEFEF}\\ 
 & & & 6 & 6 & 9\\ 
 & & & \cellcolor[HTML]{EFEFEF} 12 & \cellcolor[HTML]{EFEFEF} 12 & \cellcolor[HTML]{EFEFEF} 11 \\ 
 & && 24 & 24 & 14\\ 
 & & & \cellcolor[HTML]{EFEFEF} 48 & \cellcolor[HTML]{EFEFEF} 48 & \cellcolor[HTML]{EFEFEF} 21\\ \hline
\multicolumn{1}{l}{\multirow{5}{*}{{CV}}} & \multicolumn{1}{l}{\multirow{5}{*}{5}} & \multicolumn{1}{l|}{\multirow{5}{*}{4}} & 9 & 20 & 8\\ 
 & & & \cellcolor[HTML]{EFEFEF} 21 & \cellcolor[HTML]{EFEFEF} 35 & \cellcolor[HTML]{EFEFEF} 9\\ 
 & & & 36 & 60 & 11 \\ 
 & & & \cellcolor[HTML]{EFEFEF} 48 & \cellcolor[HTML]{EFEFEF} 80 & \cellcolor[HTML]{EFEFEF} 14\\ 
 & & & 60 & 100 & 25\\ \hline
\end{tabular}
\end{table}

The result of these benchmarks indicate that the conversion process of a canvas into a command string is negligible. Runtime appears to increase linearly as the number of atoms and connections does, but even with an impractically large model of 60 atoms and 100 arity-3 connections, the translation process on our modest workstation (a 2014 Macbook Pro) did not exceed an average of 25ms. With this is mind, we conclude that the graphical-to-textual translation process adds virtually no overhead when working on an Alloy model of a typical size., and is unlikely to be an issue for larger scale models. 

\subsection{Case Study: Debugging Real World Faulty Models}\label{sec:casestudy}
For our case study, we focus on two models from the Alloy4Fun benchmark~\cite{alloy4funbenchmark}. Alloy4Fun is a online learning platform for Alloy whose exercises have users attempt to write predicates for various models, which are checked against a back-end oracle solution. Submissions to Alloy4Fun have been anonymized and made into an open source benchmark. These models represent faulty models created by new Alloy users. While AUnit is available for any Alloy user, we envision that new users are more likely to utilize AUnit. Our case study looks to highlight how different model structures can make writing AUnit test cases tedious and error prone.

\subsubsection{Higher Arity Relations}
Often times, Alloy models consist of binary relations (2-arity). For instance, in Figure~\ref{fig:testcase}, the relation \CodeIn{header} is a binary relation of the form \CodeIn{List$\times$Node}. This is conceptually easier for a user to visualize mentally and put to paper, as the \CodeIn{header} can be envisioned a directed line that connects a single \CodeIn{List} atom to a single \CodeIn{Node} atom. However, in Alloy, it is possible for a relation of higher arity to be specified. 
To illustrate, consider the Labeled Transition System (LTS) model from the Alloy4Fun benchmark shown in Figure~\ref{fig:lts}. Line 1 introduces the signature \CodeIn{State}, which contains the relation \CodeIn{trans}. This relation is a trenary relation (3-arity) of the form \CodeIn{State$\times$Event$\times$State}. Rather than being a directed line between two atoms, \CodeIn{trans} indirectly connects two states through an intermediate \CodeIn{Event} atom. The idea of the \CodeIn{trans} relation is to that the transitions between states are triggered by events; therefore, this intermediate \CodeIn{Event} atom is an important connection between the states. 

\begin{figure}
    \centering
    \input{figures/lts}
    \caption{Faulty Model of a Labeled Transition System (LTS)}
    \label{fig:lts}
\end{figure}

For the remainder of the LTS model, line 2 introduces the \CodeIn{Init} signature as a subset (\CodeIn{in}) of the \CodeIn{State} signature, which conveys the initial state of the system, and line 3 introduces the signature \CodeIn{Event} that contains no relations itself. We elect to illustrate \tool's experience over predicate \CodeIn{inv3}, which is the third exercise in the LTS model on Alloy4Fun, as it involves the \CodeIn{trans} ternary relation in its formulation. The faulty predicate \CodeIn{inv3} (lines 6 - 8) is meant to convey that the LTS is deterministic, meaning that for every state, every \CodeIn{Event} triggers either no transition or a unique transition to a next state.  The faulty formulation uses an incorrect order of the relational joins. To illustrate, the following is the correct version of the predicate, with the difference highlighted in red for emphasis:

\begin{footnotesize}
\begin{Verbatim}[frame=lines,rulecolor=\color{lightgray}]
\Blue{all} s : State, e : Event | \Blue{lone} \Red{e}.(\Red{s}.trans)
\end{Verbatim}
\end{footnotesize}

\noindent This error is a subtle change textually, but the fault results in a formula that is trivially \textit{always} true. Namely, the faulty expression ``\CodeIn{e.trans}'' looks to form a relational join of the form \CodeIn{\underline{Event}} with \CodeIn{\underline{State}$\times$Event$\times$State}. Since there is a type mismatch, this first join will always produce an empty set. Since an empty set always satisfies the \Blue{lone} multiplicity constraint, this produces the trivially true behavior. To reveal this fault, the user needs an AUnit test case in which an \CodeIn{Event} triggers multiple possible state transitions for the same state. 

Consider the following fault revealing test case, where the red text helps highlight the behavior that the model is expected to prevent: 

\begin{footnotesize}
\begin{Verbatim}[frame=lines,rulecolor=\color{lightgray}]
\Blue{some disj}  State0, State1: State | Event0, Event1, 
Event2: Event \{\{
State = State0 + State1
trans = \Red{State1->Event0}->State0 
      + \Red{State1->Event0}->State1 
Event = Event0 + Event1 + Event2
 Init = State1
\}\}
\end{Verbatim}
\end{footnotesize}



\noindent Since \CodeIn{Event0} triggers two different transitions for \CodeIn{State1}, the valuation should not be generated, but the faulty predicate will produce it. For comparison, Figure~\ref{fig:ltscrucible} displays the same test case recreated in \tool. For the visual test case, the user can see that \CodeIn{State1} has two transitions, but both use \CodeIn{Event0}, as there are not connections drawn to \CodeIn{Event1} or \CodeIn{Event2}. In both cases, a user is likely to spot the issue with the conflict being the only values population the \CodeIn{trans} relation. 

However, consider the likely process of creating a larger test case. For example, the following is a test case that extends the previous one:

\begin{figure}
    \centering
    \includegraphics[width=0.9\columnwidth]{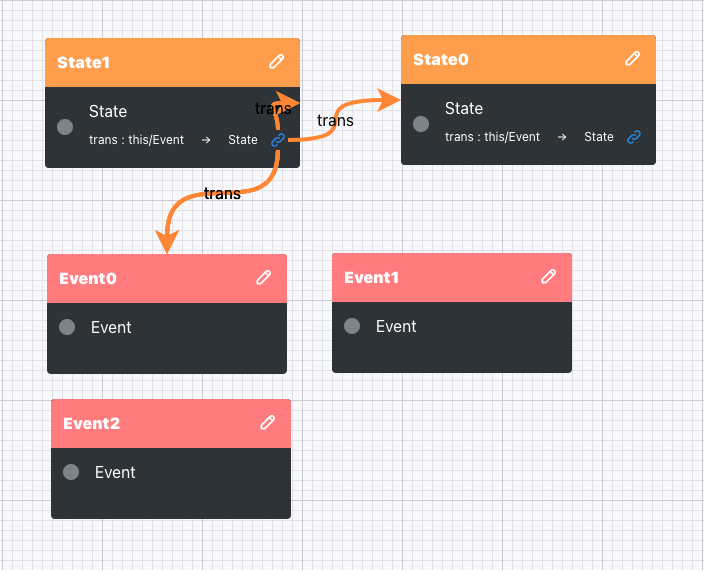}
    \caption{The faulty LTS test case in \tool}
    \label{fig:ltscrucible}
\end{figure}


\begin{footnotesize}
\begin{Verbatim}[frame=lines,rulecolor=\color{lightgray}]
\Blue{some disj}  State0, State1, State2: State | Event0, 
Event1, Event2: Event \{\{
State = State0 + State1 + State2
trans = 
    State0->Event0->State0 + State0->Event0->State1 
  + State0->Event0->State2 + State0->Event1->State0 
  + State0->Event1->State1 + State0->Event1->State2 
  + State0->Event2->State0 + State0->Event2->State1 
  + State0->Event2->State2 + State1->Event0->State0 
  + State1->Event0->State1 + State1->Event0->State2 
  + State1->Event1->State0 + State1->Event1->State1 
  + State1->Event1->State2 + State1->Event2->State0 
  + State1->Event2->State1 + State1->Event2->State2 
  + State2->Event0->State0 + State2->Event0->State1 
  + State2->Event0->State2 + State2->Event1->State0 
  + State2->Event1->State1 + State2->Event1->State2 
  + State2->Event2->State0 + State2->Event2->State1 
  + State2->Event2->State2
Event = Event0 + Event1 + Event2
 Init = State1
\}\}
\end{Verbatim}
\end{footnotesize}

\noindent which is significantly harder to follow textually. The extension to this test case is derived by using Amalgam~\cite{Amalgam} to create a maximal scenario based on the first test case. Using Amalgam allows us to highlight one of the largest, and as a result one of the more complex, fault revealing AUnit test case that a user could, in theory, create based on the current scope that preserves the same general fault revealing constraints as the small initial test case. 
There are several instances in which a users may be motivated to create larger test cases. For instance, if a user is looking to perform fault localization, repair or partial model synthesis with their AUnit test suite, past experiments reveal that larger test cases that encompass a wide degree of behavior result is notably better performance for these frameworks~\cite{AlloyFL,ARepair,ASketchABZ}.  

Realistically, if creating this test manually, the user is likely to copy, paste and then tweak assignments to the \CodeIn{trans} relation, which is an error prone process. In addition, if the user is textually specifying a relation this large, the user would realistically execute the test case and visually inspect the scenario the Analyzer produces multiple times as they build up the valuation, to make sure the valuation actually matches their expectation. 


%

\begin{figure}
    \centering
    \includegraphics[width=0.9\columnwidth]{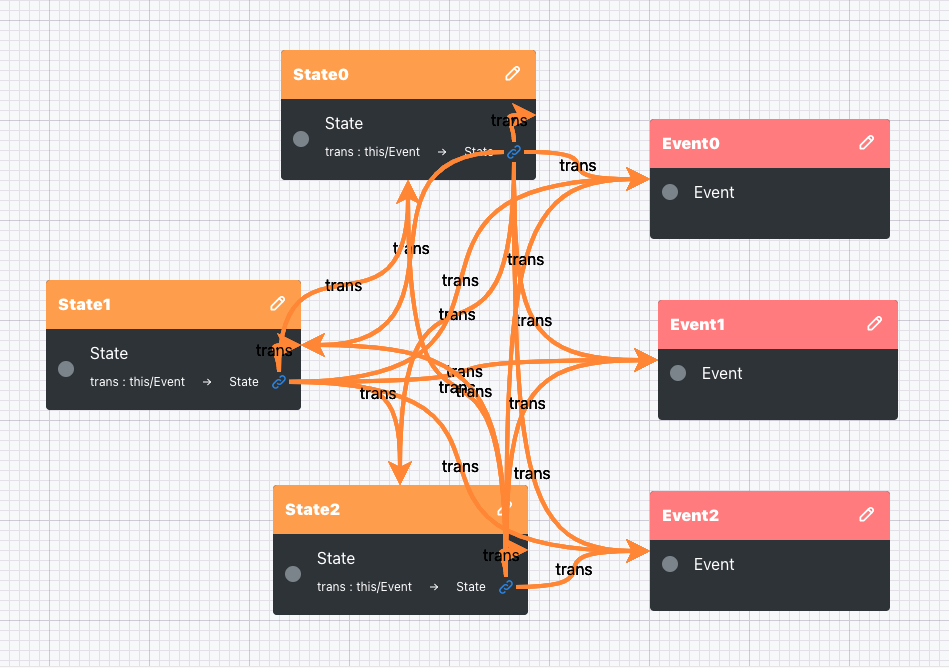}
    \caption{The maximal LTS test case in \tool}
    \label{fig:ltscrucible2}
\end{figure}

In contrast, Figure~\ref{fig:ltscrucible2} displays this same larger test case recreated in \tool. Due to the different format of creating a test, \tool{} removes the potential for copy-paste errors. More importantly, since \tool{} provides a live graphical view as the user builds up a test case, \tool{} removes the need to do repeated executions to spot-check the textual specification. These spot-checks do involve repeatedly running Alloy's backend SAT solver, although AUnit tests do not individually have a high overhead. 
While both the textual and graphical representations are cluttered, it is easier to implement lightweight interventions to make a graphical test case more readable. For instance, when a user hovers on a connection, we can gray out all unrelated atoms, easily bringing different portions of the \CodeIn{trans} relation into focus for the user. For now, users can drag and re-arrange the visual layout to better inspect connections post-creation. In contrast, there is no easy pathway to increase readability for the textual representation.

\subsubsection{Numerous Signatures and Relations}
In a recent profile of over 2000 different Alloy models~\cite{staticprofilealloy}, 
the median number of signatures and relations in a model is 8 and 2 respectively. Therefore, it is realistic to expect that a user may work with a model that contains a large number of signatures and relations. As the number of signatures and relations grow, the complexity of the valid scenarios for the models also grows. While Alloy defaults to a scope of 3 for commands, this scope is an upper bound of the size of \emph{each} signature individually, and not a collective scope. As a result, if a user has 8 signatures in their model, a valid scenario can have up to 24 atoms. In addition, the scope does not place any restrictions on the size of relations. Therefore, these 24 atoms can be interconnected in 100s of ways. All of this increases the burden for a user to mentally visualize a scenario and then textually specify the corresponding test case. In fact, it would not be surprising in this instance, if the user first drew a scenario on paper before writing the corresponding test case. 

\begin{figure}
    \centering
    \input{figures/cv}
    \caption{Faulty Model of a Curriculum Vitae Policy}
    \label{fig:cv}
\end{figure}

To illustrate how \tool{} can ease the burden of creating test cases with numerous signatures and relations, we select the \CodeIn{CV} model from the Alloy4Fun benchmark, which has 5 signatures and 4 relations, as seen in Figure~\ref{fig:cv}. Line 1 introduces an abstract signature \CodeIn{Source}. As an abstract signature, \CodeIn{Source} cannot directly have atoms itself. The next two signatures extend the \CodeIn{Source} signature. Line 2 introduces the signature \CodeIn{User}, which contains two relations: \CodeIn{profile} connects a \CodeIn{User} to any number (\CodeIn{set}) of \CodeIn{Work} elements (line 3) and \CodeIn{visible} connects a \CodeIn{User} to any number (\CodeIn{set}) of \CodeIn{Work} elements as well (line 4). Line 6 introduces the signature \CodeIn{Institution} and line 8 introduces the signature \CodeIn{Id}, neither of which define any relations. Lastly, line 9 introduces the signature \CodeIn{Work}, which contains two relations: \CodeIn{ids} connects a \CodeIn{Work} atom to at least one (\CodeIn{some}) \CodeIn{Id} atom (line 10) and \CodeIn{source} connects a \CodeIn{Work} atom to exactly one (\CodeIn{one})  \CodeIn{Source} atom (line 11). 

The faulty predicate \CodeIn{inv1} (lines 16 - 18) attempts to use subset (\CodeIn{in}) to specify that any visible work is someone's CV must be part of that person's profile. The correct version of the predicate is: 

\begin{footnotesize}
\begin{Verbatim}[frame=lines,rulecolor=\color{lightgray}]
\Blue{all} u:User | u.visible \Blue{in} u.profile 
\end{Verbatim}
\end{footnotesize}

\noindent which is similar to the incorrect formula, but constrains the subset relationship to be true for each individual person (\CodeIn{\Red{u}.visible}), rather than a universal perspective (\CodeIn{\Red{User}.visible}). As a result, for the incorrect formula, a user could have a visible work in their CV as long as at least one person has that work in their profile, even if that person is not them. Consider the following fault revealing test case: 

\begin{footnotesize}
\begin{Verbatim}[frame=lines,rulecolor=\color{lightgray}]
\Blue{some disj}  User0, User1: User | Work0, Work1, Work2: 
Work | Id0 : Id \{\{\{
   User = User0 + User1
profile = User1->Work0 + User1->Work1 + User1->Work2
visible = User0->Work0 + User0->Work1 + User0->Work2
     Id = Id0
   Work = Work0 + Work1 + Work2
    ids = Work0->Id0 + Work1->Id0 + Work2->Id0
 source = Work0->User1 + Work1->User1 + Work2->User0
\Blue{no} Institution
\}\}\}
\end{Verbatim}
\end{footnotesize}

The issue with this test case is that \CodeIn{User0} is able to have \CodeIn{Work0} visible on their CV despite not having \CodeIn{Work0} in their profile because another user (\CodeIn{User1}) has the work in their profile. In fact, for this test case, this is true for every single work that is visible on \CodeIn{User0}'s CV. At first glance, this relationship may be easy to type and confirm textually. However consider the following test case, which is an extension of the previous test case:

\begin{footnotesize}
\begin{Verbatim}[frame=lines,rulecolor=\color{lightgray}]
\Blue{some disj}  User0, User1, User2: User | Work0, Work1, 
Work2: Work | Id0, Id1, Id2: Id 
\{\{\{
   User = User0 + User1 + User2
profile = User0->Work1 + User0->Work2 + \Red{User1->Work0}
        + User1->Work1 + User1->Work2 + \Red{User2->Work0} 
        + User2->Work1 + User2->Work2
visible = \Red{User0->Work0} + User0->Work1 + User0->Work2 
        + User1->Work0 + User1->Work1 + User1->Work2 
        + User2->Work0 + User2->Work1 + User2->Work2
    Id = Id0 + Id1 + Id2
   Work = Work0 + Work1 + Work2
    ids = Work0->Id0 + Work0->Id1 + Work0->Id2 
        + Work1->Id0 + Work1->Id1 + Work1->Id2 
        + Work2->Id0 + Work2->Id1 + Work2->Id2
 source = Work0->User2 + Work1->User2 + Work2->User1
\Blue{no} Institution
\}\}\}
\end{Verbatim}
\end{footnotesize}

\begin{figure}
    \centering
    \includegraphics[width=0.9\columnwidth]{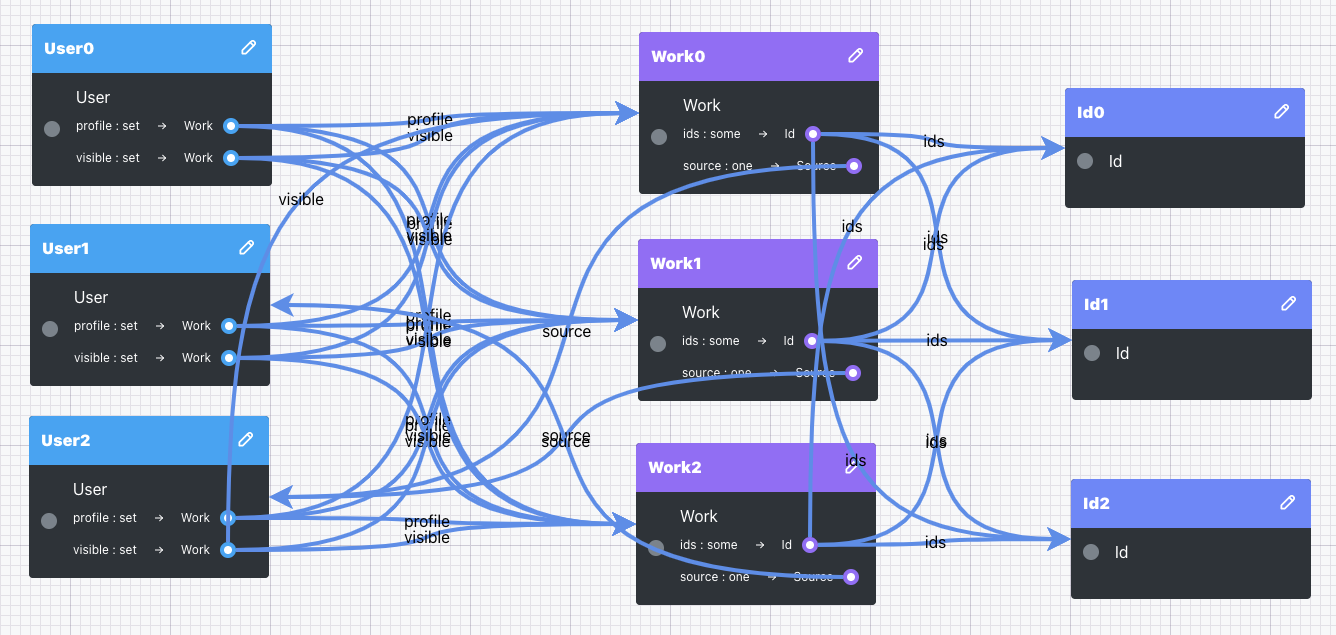}
    \caption{The Curriculum Vitae Model test in \tool{}}
    \label{fig:cv2}
\end{figure}

This test case is again derived by using Amalgam to create a maximal scenario based on the first test case~\cite{Amalgam} to also highlight one of the largest fault revealing AUnit test case a user could in theory create based on the scope. Again, the faulty behavior is revealed by the portion presented in red text. Figure~\ref{fig:cv2} displays the same test case recreated in \tool. 

As with the LTS model, the main advantage of using \tool{} is reducing the uncertainty of mentally re-creating such a long text chain for specifying a test case. In this case, rather than the majority of the complication being one relation, the complexity comes from the combination of different ways the atoms can relate to one another within the model. This still creates a high spatial cognitive burden to attempt to mentally visualize the test valuation from the text format, which is likely to result in the user incrementally writing the test and executing it to spot check that the test is written correctly.  
While the large test case one again looks cluttered, the same lightweight visual interventions mentioned earlier apply here as well, while nothing can ease the text inspection burden. 




\section{Future Work}
In \tool{}'s current form, larger test cases can become quite cluttered, as is the case for 
Figure~\ref{fig:ltscrucible2}. Although this creates some overhead for the user as they are required to track connections visually, we hold that this overhead is less than that of the alternative -- mentally visualizing a test case then writing a complex valuation textually. In future work, we will explore ways to reduce the visual clutter that a large test case creates by looking into new pathing techniques for connections and alternative visualization methods for canvases. 

In Alloy, a user can customize the Analyzer's output with a robust theming subsystem. In \tool{}'s current version, limited support for customization is available, with users having the ability to assign a custom color to each signature type. In a future release we aim to further enable the user to customize the appearance of their test with features such as the ability to rename atom instances, change the shape of the signatures, and highlight connections on hover. Enabling users to be more expressive when designing test cases will increase clarity and allow for the adoption of custom typologies within a user or organization's workflow.  

In addition, we hope to further improve our support for high arity (arity-3 and above) connections. This can be accomplished by improving the interface used to create arity-3 connections and by hardening our automated guidance techniques to ensure high arity relations have their multiplicities correctly enforced on the canvas. Optimized support for arity-3 and above connections will ensure that \tool{} is useful for the majority of Alloy models in use today.

Finally, we plan to explore how to infer the underlying model structure from a collection of graphical test cases. This would alleviate the ``how do I get started'' burden of writing software models, which a recent user study found that both novice and expert Alloy users struggle to get started writing their model~\cite{alloyformalise23study}. Specifically, based on an initial set of graphical test cases, we want to automatically create the signature paragraphs. To illustrate, from the test case in Figure~\ref{fig:testcase} (b), we can conclude that there are two signatures (\CodeIn{Node} and \CodeIn{List}) and that there are two binary relations (\CodeIn{header} to type \CodeIn{List$\times$Node} and \CodeIn{link} to type \CodeIn{Node$\times$Node}). While a single test case does not let us confirm with 100\% certainty the multiplicity of these relations, the user could supply additional tests that do. If not, we envision having an interactive process where we prod the user for clarification.

\section{Related Work}
\textbf{\textit{Testing and Debugging Techniques for Alloy.}} Cruicble aims to ease the adoption of AUnit. There are a number of testing and debugging techniques which utilize AUnit tests: $\mu$Alloy is a mutation testing framework~\cite{AUnit}, AlloyFL is a hybrid fault localization technique that uses spectrum-based and mutation-based fault localization strategies to create a ranked list of suspicious locations~\cite{AlloyFL}, and ARepair is a generate-and-valid automated repair technique that uses AUnit test cases as an oracle to evaluate potential patches~\cite{ARepair}. ICEBAR extends ARepair to consider built in Alloy assertions in addition to test cases to guide the repair~\cite{icebar}.

There are also a number of repair techniques that user built in assertions in place of AUnit tests. ATR is an Alloy repair technique that tries to find patches based on a preset number of templates and uses Alloy assertions as an oracle~\cite{ATR}. BeAFix is an automated repair technique that uses a bounded exhaustive search~\cite{BeAFix}. TAR is a mutation-oriented repair technique that is aimed at repairing Ally4Fun models, which are educational exercises~\cite{TAR}. FLACK is a fault localization technique that locates faults by using a partial max sat toolset to compare the difference between a satisfying instance of a predicate and a counterexample from an assertion over that predicate~\cite{flack}. Alloy assertions can be used to check the accuracy of predicates, but assertions need to be written correctly themselves to be beneficial.

\textbf{\textit{Drawing System Workflows.}} Our approach shares the spirit of storyboard programming, which uses user-provided graphical representations of data structures to synthesize code to perform data structure manipulations, based on the insight that it can be easier and more intuitive for a user to draw concrete data structure manipulations than to write the code~\cite{storyboardDS}. Besides, storyboard programming, there are other efforts related to drawing data structures and their transformations~\cite{ding1990framework}. \tool{} makes use of a similar insight: that it can be easier to draw examples of system behavior rather than to formally write the constraints. While not mathematical software models, there are several efforts to allow users to draw different UML diagrams~\cite{FlexiSketch,chen2003whiteboard,lank2001line}. These efforts, in particular FlexiSketch~\cite{FlexiSketch}, allow users to free hand draw portions of UML diagrams. The lessons learned from the efforts helped informed our choice of where to draw the line between free-hand drawings and a more structure drag-and-drop interface.

\section{Conclusion}
AUnit test cases give users a simple and systematic way to spot check their Alloy models for correctness. In addition, a unit testing framework helps the model development process feel closer to that of writing imperative programs for novice software modelers. However, the need to specify AUnit test cases textually is a barrier to adoption for AUnit and its supported testing infrastructures, like fault localization and automated repair. By enabling users to build AUnit test cases graphically, we bring the creation of test cases more in line with how users interact with the output of Alloy models, which is largely a graphical process. \tool{} takes this process a step further by helping guide users to create well-formed test cases based on the existing underlying model.

%
%
%
\bibliographystyle{splncs04}
\bibliography{bib}

\end{document}